# Selenium direct doping obtained high-performance-n-type $Bi_2Te_3$-based thermoelectric materials with a wide temperature range*

LIU Zhiyuan[1,2,*], MA Junjie[1], ZENG Zhaopeng[1], MA Ni[3], BA Qian[1], ZHANG Di[1], TAO Zhe[1], XIA Ailin[1,*]

1. Advanced Ceramics Research Center, School of Materials Science and Engineering, Anhui University of Technology, Ma'anshan 243002, China

2. Institute of Energy, Hefei Comprehensive National Science Center (Anhui Energy Laboratory), Hefei 230031, China

3. School of Chemistry and Materials Science, University of Science and Technology of China, Hefei 230026, China

## Abstract

$Bi_2Te_3$-based compounds have attracted great attention in near-room-temperature thermoelectric applications because of their excellent electrical transport properties and low thermal conductivity. Solid solutions and doping can effectively optimize the performance of $Bi_2Te_3$-based materials. Currently, n-type $Bi_2Te_{3-x}Se_x$ materials doped with selenium (Se) have been reported. However, the regulatory mechanisms of Se doping directly at Te sites on their defect structure, microstructure, and bandgap have not yet been systematically investigated. This work systematically investigates the regulatory behavior of Se doping directly at Te sites on the defect structure, microstructure, and bandgap of ternary n-type $Bi_2Te_{3-x}Se_x$ compounds, and its influence on thermoelectric transport properties. The Se substitution at Te sites forms n-type donor defects $Se_{Te}^{\bullet}$, inhibits the formation of Bi′Te antisite defects, and facilitates the return of Bi atoms to their intrinsic lattice sites, while introducing Te interstitial atoms ( $Te_i^{\times}$ ) and Te vacancies ( $V_{Te}^{\bullet\bullet}$ ) and optimizing both carrier concentration and mobility, thereby effectively enhancing the electrical performance. Furthermore, supersaturated Te diffuses out as interstitial atoms and precipitates to form secondary phases. Se doping enhances phonon scattering via mass and strain field fluctuations induced by point defects, leading to a significant reduction in lattice thermal conductivity. As $x$ increases, the bandgap of the sample is widened, resulting in significant suppression of the performance degradation caused by the intrinsic-excitation-induced bipolar effect. Consequently, the $Bi_2Te_{2.7}Se_{0.3}$ sample achieves a maximum average $zT$ ($zT_{ave}$) value of 0.73 in a temperature range of 300–500 K. After annealing, the optimization of the sample's microstructure leads to an enhanced power factor and reduced thermal conductivity in the $Bi_2Te_{2.4}Se_{0.6}$ sample, achieving a maximum $zT$ value of 0.81 at 420 K and a $zT_{ave}$ value of 0.76 in the temperature range of 300–500 K. These results demonstrate that Se doping directly at Te sites can broaden the temperature range corresponding to the optimal $zT$ values, and that the annealing process can further optimize the thermoelectric performance. This study provides significant insights for developing high-performance near-room-temperature thermoelectric materials applicable to broad operating temperatures.

# 1 Introduction

Thermoelectric materials serve as novel energy conversion materials capable of directly interconverting thermal and electrical energy via thermoelectric physical effects[1-3]. The comprehensive transport performance of thermoelectric materials is primarily characterized by the dimensionless figure of merit ($zT$), defined as $zT = S^2\sigma T/\kappa$. Here, $\sigma$ represents electrical conductivity, $S$ denotes the Seebeck coefficient, and $S^2\sigma$ constitutes the power factor, which reflects the comprehensive electrical transport performance. The thermal conductivity $\kappa$ indicates thermal transport performance and comprises lattice thermal conductivity $\kappa_L$ and carrier thermal conductivity $\kappa_E$. Achieving a high $zT$ value requires materials to possess a high $S^2\sigma$ and a low $\kappa$. However, these three parameters are interconnected and influenced by band structure and scattering mechanisms[4,5]. Consequently, decoupling these transport parameters is crucial for enhancing the $zT$ value.

$Bi_2Te_3$ features a narrow bandgap and a typical trigonal layered structure (rhombohedral lattice, space group $R\overline{3}m$). It forms through the periodic stacking of quintuple layers: -Te(1)-Bi-Te(2)-Bi-Te(1)-. In this structure, Te-Bi bonds are polar covalent, while the interlayer Te(1)-Te(1) interactions are governed by weak van der Waals forces[6]. The layered structure of $Bi_2Te_3$ and the weak bonding between Te(1)-Te(1) layers facilitate cleavage along the basal plane ($a$ and $b$ directions) and result in mechanical and electrical anisotropy[7]. The unique crystal structure of $Bi_2Te_3$, combined with a highly symmetric and complex band structure near the Fermi level (characterized by high carrier effective mass), yields low thermal conductivity and a large Seebeck coefficient. Furthermore, this structure allows external atoms to diffuse easily between layers, facilitating elemental doping into the $Bi_2Te_3$ lattice. Numerous studies have reported doping or co-doping $Bi_2Te_3$ with elements such as Ag[8], Na[9], Nb[10], Cu[11], Sb[12,13], Cl/W[14], and Tl/I[15]. Among these, selenium (Se)-doped $Bi_2Te_3$-based thermoelectric materials are considered optimal n-type materials for near-room-temperature applications. Due to its high solid solubility and low lattice mismatch, Se enables synergistic optimization of electrical and thermal transport. It achieves this by regulating carrier concentration at the atomic scale, enhancing phonon scattering through atomic-level defects, and maintaining band stability. This establishes Se as the premier dopant for n-type $Bi_2Te_3$-based thermoelectric materials. Extensive research has been conducted on $Bi_2Te_{3-x}Se_x$ materials with Se doping[13,16-22]. However, the mechanisms by which direct Se doping at the Te site regulates defect structures, microstructures, and bandgaps in n-type $Bi_2Te_{3-x}Se_x$ materials remain systematically unexplored. Additionally, as a narrow-bandgap semiconductor, $Bi_2Te_3$ suffers from intrinsic excitation at elevated temperatures. This triggers the bipolar effect, leading to performance degradation and severely limiting the application of $Bi_2Te_3$-based alloys across wide temperature ranges. Doping-induced band structure reconstruction can effectively widen the bandgap of $Bi_2Te_3$-based alloys[22]. This significantly suppresses the bipolar carrier concentration generated by intrinsic excitation, thereby eliminating performance attenuation caused by the bipolar effect. Consequently, this approach yields high-performance $Bi_2Te_3$-based materials with stability over a broad temperature range.

This study prepared n-type $Bi_2Te_{3-x}Se_x$ ($x$ = 0.2, 0.3, 0.4, 0.5, and 0.6) thermoelectric materials with direct Se doping at the Te site. The synthesis involved high-temperature melting, spark plasma sintering (SPS), and annealing. Samples before annealing were labeled BA02, BA03, BA04, BA05, and BA06, while those after annealing were labeled AA02, AA03, AA04, AA05, and AA06. We systematically investigated the effects of direct Se doping at the Te site on the phase composition, crystal defects, microstructure, and bandgap of $Bi_2Te_{3-x}Se_x$-based materials. Furthermore, we explored the influence of doping on electrical and thermal transport properties. Appropriate Se doping regulates carrier concentration and mobility by adjusting internal defect structures, such as $Se_{Te}^{\bullet}$ donor defects, $Bi_{Te}^{'}$ antisite defects, and $V_{Te}^{\bullet\bullet}$ vacancy defects. This effectively enhances electrical transport performance. Simultaneously, Se doping-induced band structure reconstruction widens the bandgap. This significantly reduces the concentration of thermally excited bipolar carriers, effectively suppressing performance degradation caused by the bipolar diffusion effect from intrinsic excitation. As a result, the temperature range corresponding to optimal performance is broadened. Additionally, annealing further optimizes the microstructure, leading to enhanced thermoelectric performance. This study systematically examines how direct doping regulates defect structures, microstructures, and bandgaps in n-type $Bi_2Te_3$-based materials. It clarifies the intrinsic relationships with thermoelectric transport optimization and deepens the understanding of point defects and carrier transport behavior. These findings provide insights for developing high-performance, near-room-temperature thermoelectric materials suitable for wide temperature ranges.

# 2 Experimental Methods

## 2.1 Material Preparation

High-purity Bi powder (99.99%), Te powder (99.99%), and Se powder (99.99%) were weighed according to the nominal composition $Bi_2Te_{3-x}Se_x$ (BTS) ($x$ = 0.2-0.6). Using a Partulab MRVS-1002 system, these high-purity powders were mixed and sealed in quartz tubes under a vacuum of less than 0.1 MPa. The sealed tubes were transferred to a programmable furnace, heated to 1073 K over 5 h, held at this temperature for 10 h, and then naturally cooled to room temperature inside the furnace. The resulting ingots were ultrasonically cleaned, ground, and sieved to obtain fine powders. The powders were loaded into graphite molds and sintered at 723 K for 5 min using an SPS apparatus (SPS-1050, Dr. Sinter) to produce cylindrical bulk materials with a diameter of 15 mm. Finally, the samples were sealed in quartz tubes and annealed in a programmable furnace at 573 K for 72 h to obtain the annealed samples.

## 2.2 Characterization and Performance Testing

The phase composition of the samples was analyzed using powder X-ray diffraction (XRD, Bruker D8 ADVANCE; Cu $K\alpha$ radiation source, $\lambda$ = 0.15418 nm). Room-temperature Raman spectroscopy was performed using a laser confocal micro-Raman spectrometer (inVia, Renishaw, UK). The microstructure and composition of the samples were characterized using a field-emission scanning electron microscope (NANO SEM430, FEI) and energy-dispersive X-ray spectroscopy (EDS), respectively. The UV-Vis absorption spectra of the samples were measured in the wavelength range of 200-800 nm using a Shimadzu UV3600 instrument to determine the bandgap. The electrical resistivity $\rho$ and Seebeck coefficient $S$ were measured using the four-probe method with a Linseis system (LSR-3, Germany) under high-purity helium (He) atmosphere. The thermal conductivity $\kappa$ was calculated using the formula $\kappa = \lambda\rho_0C_\text{p}$, based on the measured thermal diffusivity ($\lambda$), specific heat capacity ($C_\text{p}$), and sample density ($\rho_0$). The thermal diffusivity $\lambda$ and specific heat capacity $C_\text{p}$ were measured using a laser flash thermal conductivity tester (LFA 457, Netzsch, Germany), while the density $\rho_0$ was determined using the Archimedes method. The electronic thermal conductivity $\kappa_\text{E}$ was calculated using the Wiedemann-Franz law: $\kappa_\text{E} = L\sigma T$, where $L$ is the Lorenz number. The lattice thermal conductivity $\kappa_\text{L}$ was obtained using the formula $\kappa_\text{L} = \kappa - \kappa_\text{E}$. The uncertainties for $\sigma$ and $\kappa$ were ±(5%-7%), while those for $S$ and $zT$ were ±5%. Room-temperature Hall measurements were conducted using a Lake Shore 8400 system. The Hall coefficient $R_H$ and electrical resistivity $\rho$ were measured at 300 K under a reversible magnetic field of 1.0 T. The carrier concentration ($n$) and mobility ($\mu_H$) were calculated using the measured resistivity $\rho$ and Hall coefficient $R_H$ via the formulas $n = 1/(R_H e)$ and $\mu_H = R_H/\rho$, where $e$ represents the elementary charge.

# 3 Results and Discussion

## 3.1 Phase Composition and Microstructure

Figure 1(a) presents the XRD patterns of $Bi_2Te_{3-x}Se_x$ ($x$ = 0.2-0.6) powder samples. The characteristic diffraction peaks of all samples align with the standard card JCPDS 85-0439 for rhombohedral $Bi_2Te_3$, and no characteristic peaks of other impurity phases are observed. This indicates that the prepared $Bi_2Te_{3-x}$ $Se_x$ series samples maintain a single-phase structure, and Se doping does not induce impurity phase formation. As shown in Figure 1(b), with increasing Se doping content, the (0 1 5) diffraction peak shifts to the right within the diffraction angle range of 26°-29°. The magnitude of this shift increases with higher doping levels. This phenomenon occurs because the atomic radius of Se (~1.16 Å) is smaller than that of Te (~1.36 Å) in the Se-doped $Bi_2Te_3$ system. Consequently, the $Bi_2Te_{3-x}Se_x$ lattice contracts as $x$ increases. According to Bragg's law, $2d\sin q = \lambda$, a decrease in interplanar spacing $d$ leads to an increase in the sine of the diffraction angle $\theta$ under a fixed wavelength $\lambda$. Therefore, the shift of diffraction peaks to higher angles with increased Se doping confirms the successful substitution of Te by Se in the $Bi_2Te_3$ lattice. Figure 1(c) displays the XRD patterns of the annealed $Bi_2Te_{3-x}Se_x$ powder samples. The characteristic peaks of the annealed samples remain consistent with the standard card for rhombohedral $Bi_2Te_3$, and no impurity phase peaks are observed.

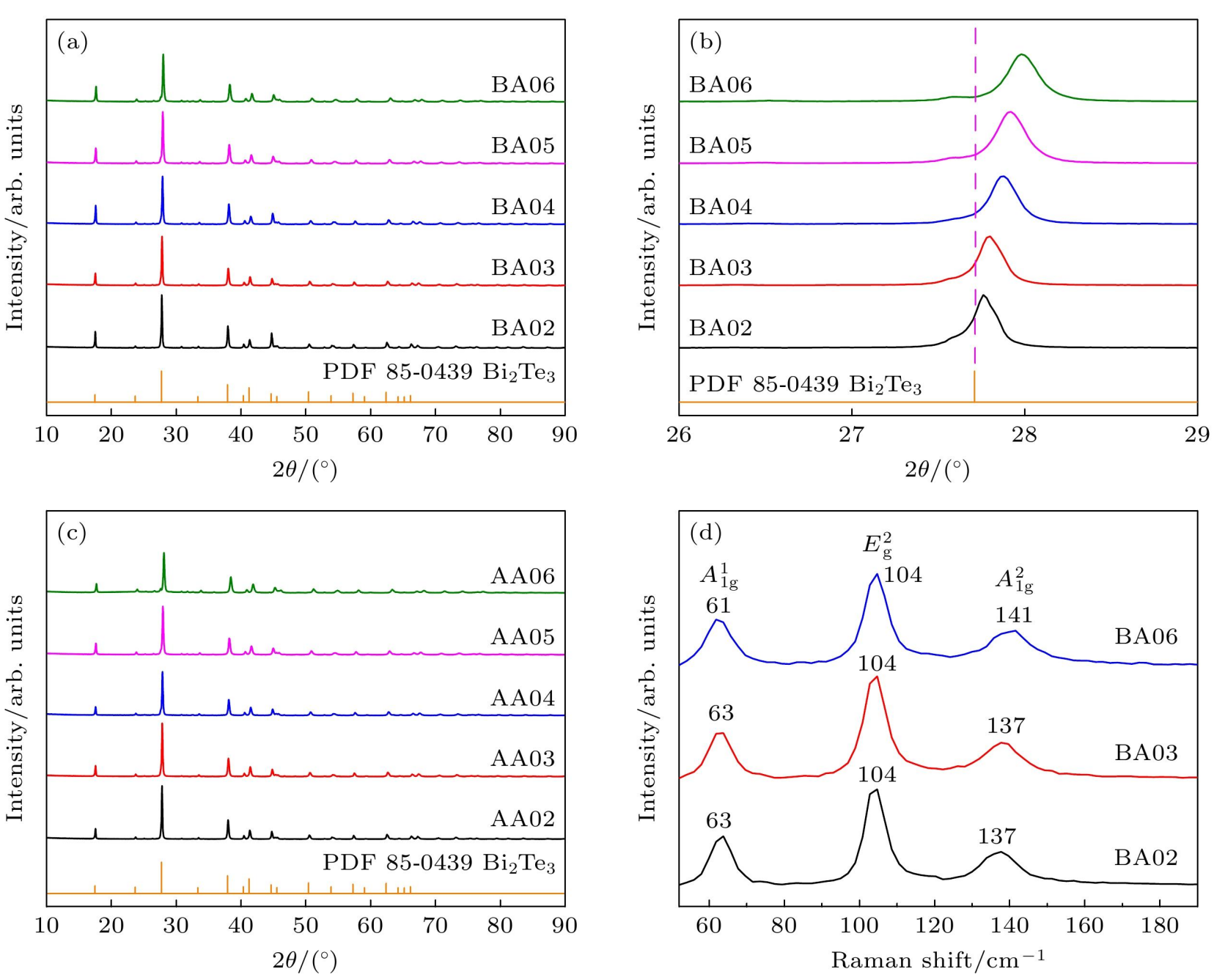

**Fig. 1** (a) XRD patterns of $Bi_2Te_{3-x}Se_x$ samples before annealing; (b) zoomed XRD patterns of samples before annealing in the 2$\theta$ range of 26°–29°; (c) XRD patterns of $Bi_2Te_{3-x}Se_x$ samples after annealing; (d) Raman spectra of $Bi_2Te_{3-x}Se_x$ samples before annealing.

To investigate the impact of direct Se doping on the crystal structure of the $Bi_2Te_3$ matrix, we measured the Raman spectra of the samples, as shown in Figure 1(d). $Bi_2Te_3$ forms through the periodic stacking of quintuple layers, specifically Te(1)-Bi-Te(2)-Bi-Te(1). These layers bind via van der Waals forces, where atoms undergo thermally induced vibrations (lattice vibrations) at their lattice positions. Figure 1(d) reveals three primary Raman optical phonon peaks for all samples: $A_{1g}^{(1)}$ (63 cm$^{-1}$), $E_g^{(2)}$ (104 cm$^{-1}$), and $A_{1g}^{(2)}$ (137 cm$^{-1}$). The $A_{1g}^{(1)}$ and $A_{1g}^{(2)}$ modes correspond to non-degenerate symmetric vibrations that produce out-of-plane motion. They reflect the in-phase vibration mode of Bi and $Te^{(1)}$ atoms and the out-of-phase vibration mode of Bi and $Te^{(1)}$ atoms, respectively. The $E_g^{(2)}$ mode corresponds to doubly degenerate symmetric vibrations that produce in-plane motion, reflecting the vibration modes between Bi atoms and between $Te^{(1)}$ atoms[23,24]. As the Se doping level increases, the $A_{1g}^{(2)}$ Raman peak of the $x$ = 0.6 sample shifts to the right. This phenomenon likely occurs because Se, which has higher electronegativity, replaces Te atoms. This substitution shortens the chemical bonds between Bi and $Te^{(1)}$, thereby shifting the vibration to higher frequencies[25].

Figures 2(a)-(f) present typical field emission scanning electron microscopy (FESEM) images of the fracture surfaces of $Bi_2Te_{3-x}Se_x$ ($x$ = 0.2, 0.3, 0.6) samples

before and after annealing. All samples exhibit the characteristic layered structure of $Bi_2Te_3$-based materials. Backscattered electron images (BEIs) of the polished surfaces of the $Bi_2Te_{3-x}Se_x$ samples before annealing reveal minor impurity phases alongside the primary $Bi_2Te_3$ phase, as shown in Figures 2(g)-(i). Energy-dispersive X-ray spectroscopy (EDS) analysis identifies this impurity phase as Te, as illustrated in the inset of Figure 2(h). This finding appears inconsistent with the X-ray diffraction (XRD) results. The discrepancy arises because the Te impurity content in the samples falls below the XRD detection limit (less than 1%). Consequently, characteristic Te diffraction peaks do not appear in the XRD patterns but are detectable in the BEI images of localized magnified areas on the polished surfaces. The heterogeneous interfaces formed by the sparse distribution of the Te phase may enhance phonon scattering, contributing to the reduction of lattice thermal conductivity in the matrix[26]. Additionally, potential barrier scattering at these heterogeneous interfaces may optimize the electrical transport properties of the material. Figures 2(g)-(i) indicate that the content of the Te impurity phase increases with higher doping levels. Based on theories of crystal growth thermodynamics, kinetics, and defect chemistry[27-30], the increase in Te impurity content with higher Se doping may stem from the following mechanism: In the $Bi_2Te_3$ crystal, Te occupies anion sites (valence -2), while Bi occupies cation sites (valence +3). When Se (valence -2) substitutes for Te sites, it forms $Se_{Te}^{\times}$. Upon ionization, $Se_{Te}^{\times} \rightarrow Se_{Te}^{\bullet} + e^{'}$ releases electrons into the conduction band, acting as an n-type donor defect ($Se_{Te}^{\times} \rightarrow Se_{Te}^{\bullet} + e^{'}$). Furthermore, Se substitution for Te may induce Te atom supersaturation. As Se occupies Te sites, the displaced Te atoms are expelled from the lattice and temporarily stored as $Te_2$ gas for subsequent reactions. The possible defect equation for this process is

$$Se + Te_{Te}^{\times} \rightarrow Se_{Te}^{\bullet} + e^{'} + \frac{1}{2}Te_2(g), \tag{1}$$

Here, $Te_{Te}^{\times}$ represents Te occupying a normal lattice site, $Se_{Te}^{\bullet}$ denotes a positively charged donor defect, $e^{'}$ signifies a negatively charged electron, and $Te_2(g)$ refers to $Te_2$ gas expelled from the lattice. When Se occupies a Te site ($Se_{Te}$), the antisite defect $Bi_{Te}^{'}$ is suppressed. Excess Se drives Bi from antisite positions back to intrinsic sites. This process simultaneously releases Te interstitial atoms ( $Te_i^{\times}$ ) and forms Te vacancies ($V_{Te}^{\bullet\bullet}$). The Te interstitial atoms diffuse to grain boundaries and aggregate, forming elemental Te precipitates. Based on defect Eq. (1), the potential defect reaction for this process is

$$Bi_{Te}^{'} + \frac{1}{2}Te_2(g) \rightarrow Bi_{Bi}^{\times} + Te_i^{\times} + V_{Te}^{\bullet\bullet} + 3e^{'}, \tag{2}$$

$$Te_i^{\times} \rightarrow Te_{n(precipitate).} \tag{3}$$

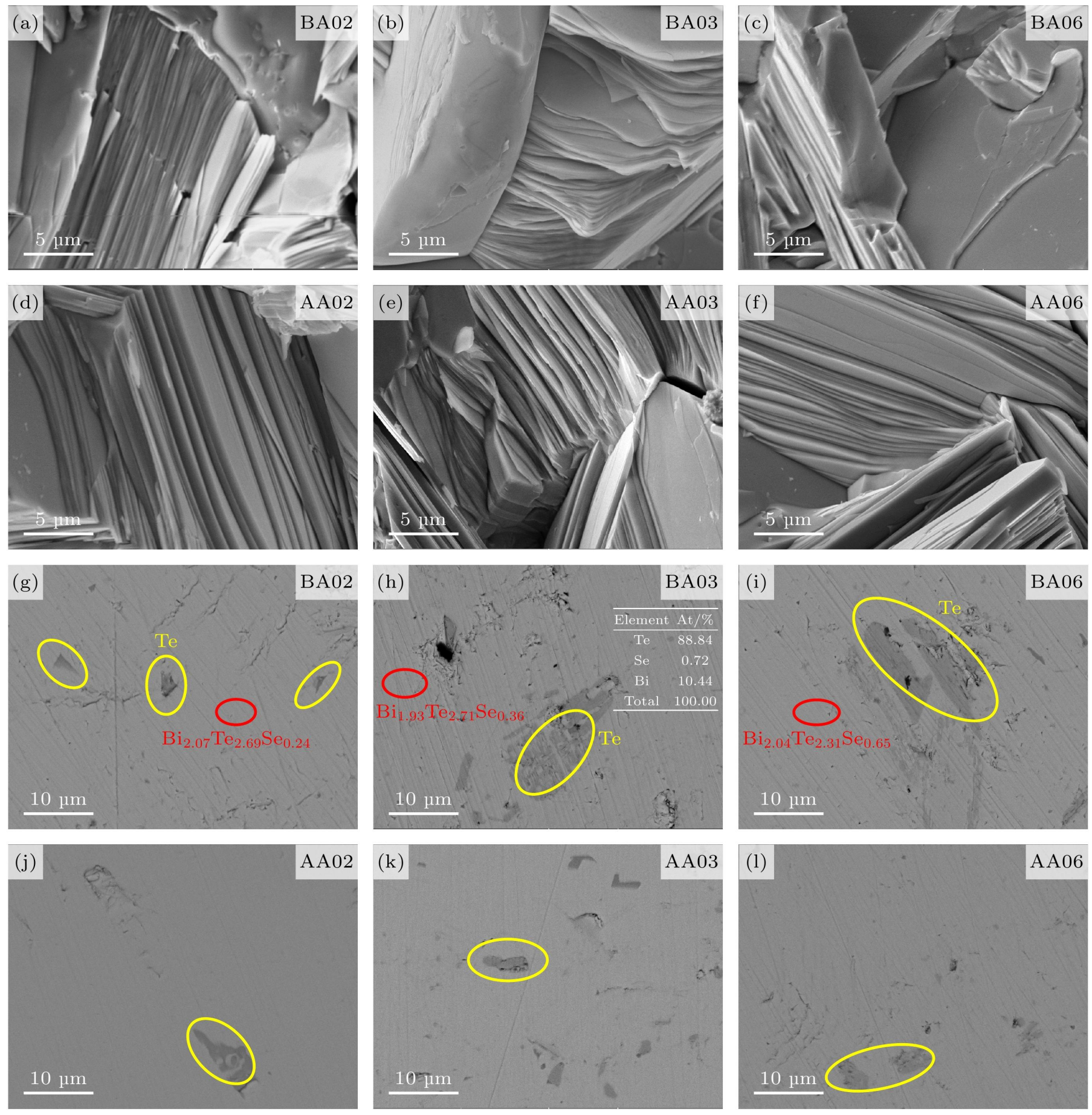

**Fig. 2** Typical FESEM images of fracture surfaces of the $Bi_2Te_{3-x}Se_x$ samples (a)–(c) before and (d)–(f) after annealing; (g)–(i) backscattered electron images (BEIs) of the polished surface of the $Bi_2Te_{3-x}Se_x$ samples before annealing, the inset shows the analysis results of the EDS spectrum in the yellow circle area in panel (h); (j)–(l) BEIs of the polished surface of the $Bi_2Te_{3-x}Se_x$ samples after annealing.

FESEM cross-sectional images of $Bi_2Te_{3-x}Se_x$ samples before and after annealing reveal that annealing does not alter the sample morphology. The material retains its typical layered structure, although grain growth occurs slightly after annealing. Such grain growth facilitates carrier migration, thereby enhancing electrical transport properties[31]. Furthermore, a significant reduction in the Te impurity phase is evident after annealing (Fig. 2(j)-(l)). This reduction may stem from two factors: 1) slight volatilization of Te during prolonged annealing; and 2) thermally activated diffusion of Te atoms, which promotes their dissolution into the matrix lattice or annihilation of defects, thereby reducing Te-rich impurity phases. Secondary electron images (SEIs) of the polished sample surfaces show that Te volatilization leaves a small number of pores, as illustrated in Fig. 3(a)-(d). The reduction of secondary phases and the presence of these sparse pores may help improve the thermoelectric performance of the samples. EDS elemental mapping of the polished surfaces before (Fig. 3(e)) and after (Fig. 3(f)) annealing indicates a more uniform elemental distribution in the annealed material. Additionally, crystallinity calculations based on XRD data demonstrate an increase in crystallinity after annealing compared to the pre-annealed state (Table 1). This improvement likely results from the slight grain growth and the reduction of the Te impurity phase induced by annealing (Fig. 2). The uniform elemental distribution and enhanced crystallinity are expected to optimize the transport properties of the material.

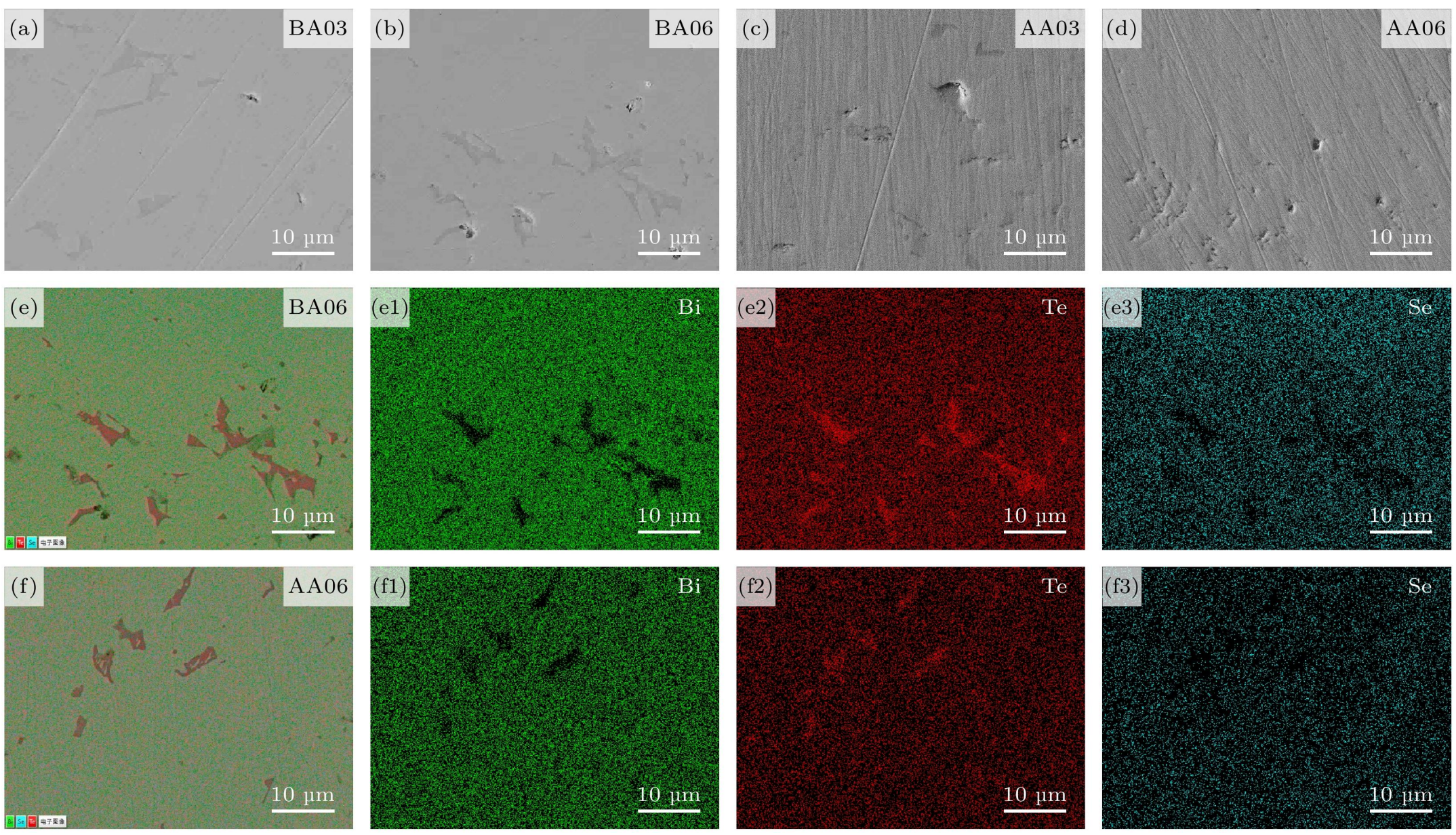

**Fig. 3** (a)–(d) Typical SEI images of the polished surface of samples BA03, BA06, AA03, AA06; (e), (f) BEI images of the polished surface of BA06 and AA06 samples, along with EDS mapping of Bi, Te, and Se elements.

**Table 1** EDS composition, room temperature Hall transport parameters (Hall coefficient $R_H$, carrier concentration $n$ and mobility $\mu_H$), density-of-states, effective mass of $Bi_2Te_{3-x}Se_x$ samples before annealing, as well as crystallinity of $Bi_2Te_{3-x}Se_x$ samples before and after annealing.

| Sample | EDS Composition | $R_H$ ($cm^3 \cdot C^{-1}$) | n ($10^{19}$ $cm^{-3}$) | $\mu_H$ ($cm^2 \cdot V^{-1} \cdot s^{-1}$) | $m^*/m_e$ | Crystallinity/% (Before/After Annealing) |
|---|---|---|---|---|---|---|
| $x = 0.2$ | $Bi_{2.07}Te_{2.69}Se_{0.24}$ | 1.28 | 4.87 | 128.82 | 1.44 | 80.79/87.13 |
| $x = 0.3$ | $Bi_{1.93}Te_{2.71}Se_{0.36}$ | 1.18 | 5.29 | 141.36 | 1.32 | 82.30/86.46 |
| $x = 0.6$ | $Bi_{2.04}Te_{2.31}Se_{0.65}$ | 1.37 | 4.56 | 146.61 | 1.15 | 88.82/92.05 |

## 3.2 Electrical Transport Properties

Table 1 presents the Hall transport parameters of $Bi_2Te_{3-x}Se_x$ samples at room temperature. All samples exhibit negative Hall coefficient ($R_H$) values, indicating n-type conduction characteristics. We calculated the carrier concentration ($n$) and Hall mobility ($\mu_H$) of the $Bi_2Te_{3-x}Se_x$ samples at room temperature by averaging multiple Hall measurements. As the Se doping level increases, the $R_H$ of the samples initially decreases and then increases, $\mu_H$ increases, whereas $n$ initially increases and then decreases. The electrons in the $Bi_2Te_{3-x}Se_x$ samples originate primarily from two sources. First, intrinsic anion vacancies $V_{Te}^{\bullet\bullet}$ formed during crystal growth generate electrons[32]. Second, donor-like effects during the grinding and sintering processes produce electrons[31,33]. The defect equation for the donor-like effect is expressed as[31]

$$2V_{Bi}^{'''} + 3V_{Te}^{\bullet\bullet} + Bi_{Te}^{'} \rightarrow V_{Bi}^{'''} + 2Bi_{Bi}^{\times} + 4V_{Te}^{\bullet\bullet} + 6e^{'}. \tag{4}$$

Consequently, the electron concentration generated by donor-like effects is closely related to the concentration of antisite defects $Bi_{Te}^{'}$. The formation of $Bi_{Te}^{'}$ depends on the electronegativity and atomic radius differences between elements, as well as anion vacancies $V_{Te}^{\bullet\bullet}$. Smaller differences in electronegativity and atomic radius lead to lower antisite defect formation energies, facilitating defect formation. The Bi-Se pair exhibits a larger electronegativity difference than the Bi-Te pair ($\chi_{Bi} = 2.02$, $\chi_{Te} = 2.1$, $\chi_{Se} = 2.55$). Under Se doping, the formation energy of antisite defects increases, which hinders their formation[34,35]. As the Se doping level increases, $Se_{Bi}^{'}$ antisite defects are gradually suppressed. Since $V_{Te}^{\bullet\bullet}$ acts as an acceptor defect that compensates for n-type carriers (increasing hole concentration), its suppression favors an increase in electron concentration. Furthermore, as previously mentioned, Se occupying Te sites suppresses $Bi_{Te}^{'}$ antisite defects. Excess Se drives Bi from antisites back to intrinsic sites, simultaneously releasing Te interstitial atoms and forming anion vacancies $V_{Te}^{\bullet\bullet}$. These vacancies act as donor defects (providing two electrons), thereby increasing the carrier concentration of the material, as shown in Eq. (2). Therefore, when $x \leq 0.3$, the increase in carrier concentration primarily stems from the suppression of $Bi_{Te}^{'}$ antisite defects and the generation of anion vacancies $V_{Te}^{\bullet\bullet}$. When $x > 0.3$, the carrier concentration decreases, likely due to three reasons: 1) At high Se doping levels, Te sites are heavily occupied by Se, saturating the number of formable $V_{Te}^{\bullet\bullet}$; the electron contribution reaches an upper limit and then declines. 2) Excess Se begins to occupy Bi vacancies, forming acceptor defects $Se_{Bi}^{'}$ that neutralize electrons, leading to a decrease in electron concentration. 3) An increase in Te impurity phases hinders carrier transport, reducing both carrier concentration and mobility. Additionally, Te impurity phases influence the carrier concentration of the samples. At low Se doping levels ($x \leq 0.3$), Te impurity phases are scarce, and carrier concentration increases with Se doping. This is mainly attributed to the suppression of $Bi_{Te}^{'}$ antisite defects and the potential formation of anion Te vacancies $V_{Te}^{\bullet\bullet}$. At high Se doping levels ($x > 0.3$), Te impurity phases increase, and carrier concentration decreases with higher Se content. This is ascribed to the self-compensation effect where acceptor defects $Se_{Bi}^{'}$, formed by excess Se occupying Bi vacancies, neutralize electrons, along with the hindrance of carrier transport by impurity phases, which reduces the effective carrier concentration.

When $x \leq 0.3$, $\mu_H$ increases significantly with rising Se content, due to four reasons: 1) Minimization of lattice strain. The atomic radius of Se is smaller than that of Te. At low Se doping levels, the substitution of Se for Te is moderate, keeping lattice contraction controllable and distortion energy below the critical threshold, which significantly reduces the probability of phonon-electron scattering[36]. 2) Weakened ionized impurity scattering. At low Se doping levels, defects are predominantly $Se_{Te}^{'}$ with moderate concentration, avoiding scattering superposition caused by multiple defect types, resulting in low ionized impurity scattering rates. 3) At low Se doping levels, antisite defects such as $Bi_{Te}^{'}$ (acceptor defects) are suppressed due to high formation energies, preventing enhanced scattering caused by the mutual cancellation of donor and acceptor defects. 4) Increased Se doping improves sample crystallinity (Table 1), which also facilitates carrier transport. When $x > 0.3$, the rate of increase in $\mu_H$ slows down as Se doping increases. Higher Se doping leads to increased concentrations of Te vacancies ($V_{Te}^{\bullet\bullet}$) and Te interstitial atoms ( $Te_i^{\bullet\bullet}$ ). These defects scatter carriers, reducing the rate of mobility increase. Furthermore, the increase in Te impurity phases hinders carrier transport, causing a decline in the mobility of some carriers.

Figures 4(a)-(c) show the temperature dependence of electrical conductivity $\sigma$, Seebeck coefficient $S$, and power factor $S^2\sigma$ for as-annealed $Bi_2Te_{3-x}Se_x$ samples in the range of 300–500 K. The results indicate that within the 300–500 K temperature range, the electrical conductivity $\sigma$ of the samples decreases monotonically with increasing temperature, exhibiting typical metallic conduction behavior characteristic of heavily doped semiconductors. Meanwhile, $\sigma$ first increases and then decreases with increasing Se doping (Fig. 4(a)). This trend corresponds to the carrier concentration, which first increases and then decreases with higher Se doping (Table 1). At 300 K, the BA04 sample exhibits the maximum $\sigma$ of $13.33\times10^4$ S/m. As shown in Fig. 4(b), in the 300–500 K temperature range, the absolute value of $S$ for samples with low Se doping ($x \leq 0.3$) first increases and then decreases significantly with rising temperature. The significant decrease is caused by bipolar conduction induced by intrinsic excitation. For samples with high Se doping ($x > 0.3$), the absolute value of $S$ first increases and then decreases slightly with rising temperature. Compared to low-doping samples, the high-temperature $S$ of samples with $x > 0.3$ does not show a significant decrease, which is attributed to the effective suppression of intrinsic excitation. In the low-temperature region, $S$ first decreases and then increases with higher Se doping, showing a trend opposite to that of $\sigma$. In the high-temperature region, the $S$ of samples with $x > 0.3$ is significantly higher than that of the BA02 sample, stemming from the marked suppression of high-temperature intrinsic excitation in heavily doped samples.

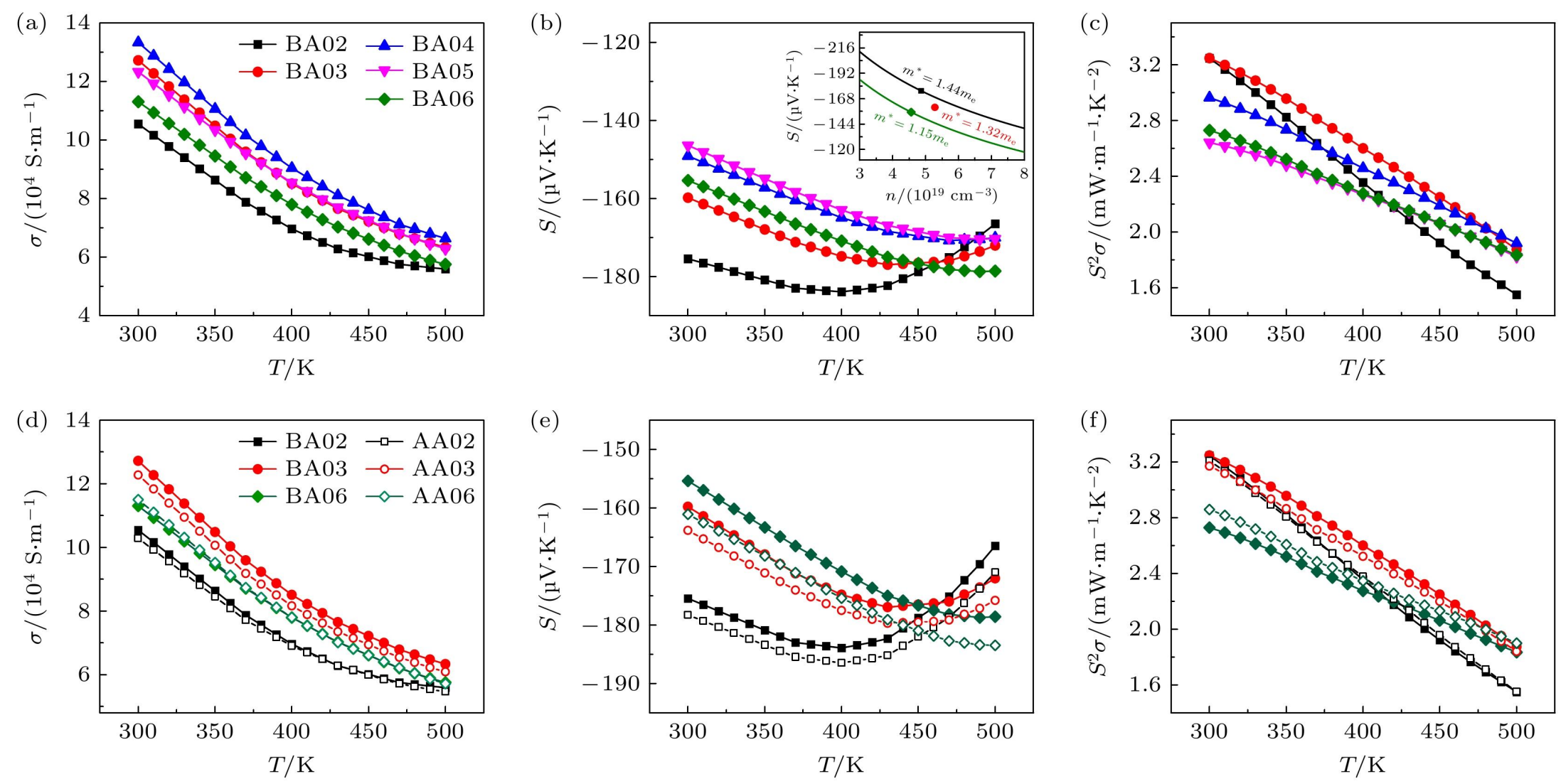

**Fig. 4** Temperature dependences of (a) electrical conductivity, (b) Seebeck coefficient and (c) power factor of $Bi_2Te_{3-x}Se_x$ samples before annealing in the temperature range of 300–500 K. The inset in panel (b) shows the $S$-$n$ curves at 300 K achieved by the Pisarenko relation with $m^*$= 1.15$m_e$ and $m^*$= 1.44$m_e$. Temperature dependences of (d) electrical conductivity, (e) Seebeck coefficient and (f) power factor of $Bi_2Te_{3-x}Se_x$ ($x$ = 0.2, 0.3, and 0.6) samples before and after annealing in the temperature range of 300–500 K.

To elucidate the mechanism underlying the variation in $S$, we employed the single parabolic band (SPB) model. This approach utilizes the energy-dependent relaxation time and degenerate approximation to analyze changes in $S$[37-39]. Eq. (5) expresses $S$[40]. This theoretical framework facilitates a systematic analysis of the physical essence governing $S$ parameter variations during electrical transport:

$$S = \frac{8\pi^2 k_B^2}{3eh^2} m^* T \left(\frac{\pi}{3n}\right)^{2/3}, \tag{5}$$

Here, $k_B$ denotes the Boltzmann constant, $h$ represents the Planck constant, $m^*$ indicates the density of states (DOS) effective mass, and $T$ signifies the absolute temperature. The inset in Figure 4(b) illustrates the functional relationship between $S$ and $n$ for $Bi_2Te_{3-x}Se_x$ samples at 300 K, derived from the single parabolic band (SPB) model. We calculated $m^*$ using Eqs. (6) and (7):

$$S = \pm \frac{k_B}{e}\left[\eta - \frac{(r+5/2)F_{r+3/2}(\eta)}{(r+3/2)F_{r+1/2}(\eta)}\right], \tag{6}$$

$$n = \frac{4\pi(2m^* k_B T)^{3/2}}{h^3} F_{1/2}(\eta). \tag{7}$$

Here, $F_i(\eta)$ denotes the Fermi-Dirac integral,

$$F_i(\eta) = \int_0^\infty \frac{\chi^i}{1+e^{\chi-\eta}} d\chi; \tag{8}$$

$\eta$ denotes the reduced Fermi level; $r$ represents the scattering factor; $\chi = E/(k_B T)$ is the normalized energy variable in the Fermi integral; and $i$ is a parameter indicating the order of integration. The calculated effective mass $m^*$ for $Bi_2Te_{3-x}Se_x$ ($x$ = 0.2, 0.3, 0.6) samples ranges from $1.15m_e$ to $1.44m_e$ (Table 1), where $m_e$ is the free carrier mass. In the inset of Figure 4(b), the Pisarenko relationship for the BA02 sample is represented by the black line. If changes in the Seebeck coefficient $S$ were solely dependent on the carrier concentration $n$, the $S$ values of other doped samples should lie on this curve. However, the $S$ values of other doped samples deviate from the black curve, indicating that changes in $S$ depend not only on $n$ but also on $m^*$. According to Equation (4), $S$ is proportional to $m^*$ and inversely proportional to $n$. As observed in the inset of Figure 4(b), $S$ decreases with increasing Se doping content. This results from the combined effects of a decrease in the effective mass $m^*$ and variations in $n$ (which initially increases and then decreases).

Figure 4(c) illustrates the temperature dependence of the power factor $S^2\sigma$ for $Bi_2Te_{3-x}Se_x$ samples before annealing within the 300–500 K temperature range. As the temperature rises, the $S^2\sigma$ of all samples gradually decreases. In the 300–450 K range, the BA03 sample exhibits a higher $S^2\sigma$ than the other samples, attributable to its higher electrical conductivity $\sigma$ and moderate $S$. The BA03 sample achieves a maximum $S^2\sigma$ of 3.25 mW/(K$^2$·m) at 300 K. Notably, in the high-temperature region (425–500 K), the $S^2\sigma$ of samples with $x \geq 0.3$ is significantly greater than that of the BA02 sample. This is attributed to the significant suppression of intrinsic excitation in the former at high temperatures, resulting in substantially higher $\sigma$ and $S$ values compared to the BA02 sample. These results indicate that appropriate Se doping can regulate carrier transport and optimize the electrical transport properties of $Bi_2Te_{3-x}Se_x$-based materials by effectively suppressing high-temperature intrinsic excitation.

Figures 4(d)-(f) display the temperature dependence of electrical transport parameters ($\sigma$, $S$, and $S^2\sigma$) for $Bi_2Te_{3-x}Se_x$ samples before and after annealing in the 300-500 K range. Compared to the pre-annealing state, the temperature dependence trends of these electrical transport parameters remain largely unchanged after annealing. The electrical conductivity $\sigma$ shows minimal variation before and after annealing (Figure 4(d)), with only a slight decrease observed for the $x$ = 0.3 sample. This may result from carrier scattering caused by the formation of minor micropores after annealing (Figure 3). The Seebeck coefficient $S$ increases for all samples after annealing (Figure 4(e)), which correlates with the slight decrease in $\sigma$. Notably, for the heavily doped sample ($x$ = 0.6), $\sigma$ remains relatively stable after annealing, whereas $S$ increases significantly. This may be attributed to the reduction of secondary phases and the energy filtering effect on carriers caused by micropores formed during annealing (Figure 3). For lightly doped samples ($x \leq 0.3$), the $S^2\sigma$ shows little change

due to the decrease in $\sigma$ and the slight increase in $S$ after annealing; specifically, the $S^2\sigma$ of the $x$ = 0.3 sample decreases only slightly. In contrast to the lightly doped samples, the $S^2\sigma$ of the heavily doped sample ($x$ = 0.6) improves after annealing (Figure 4(f)) due to the significant increase in its $S$.

### 3.3 Thermal Transport Properties

Figure 5 presents the thermal transport properties of $Bi_2Te_{3-x}Se_x$ samples before annealing within the 300–500 K temperature range. According to the Wiedemann-Franz law, the electronic thermal conductivity $\kappa_E$ can be calculated using $\kappa_E = L\sigma T$. The Lorenz number $L$ is determined using the Single Parabolic Band (SPB) model via Equation (9)[41,42]:

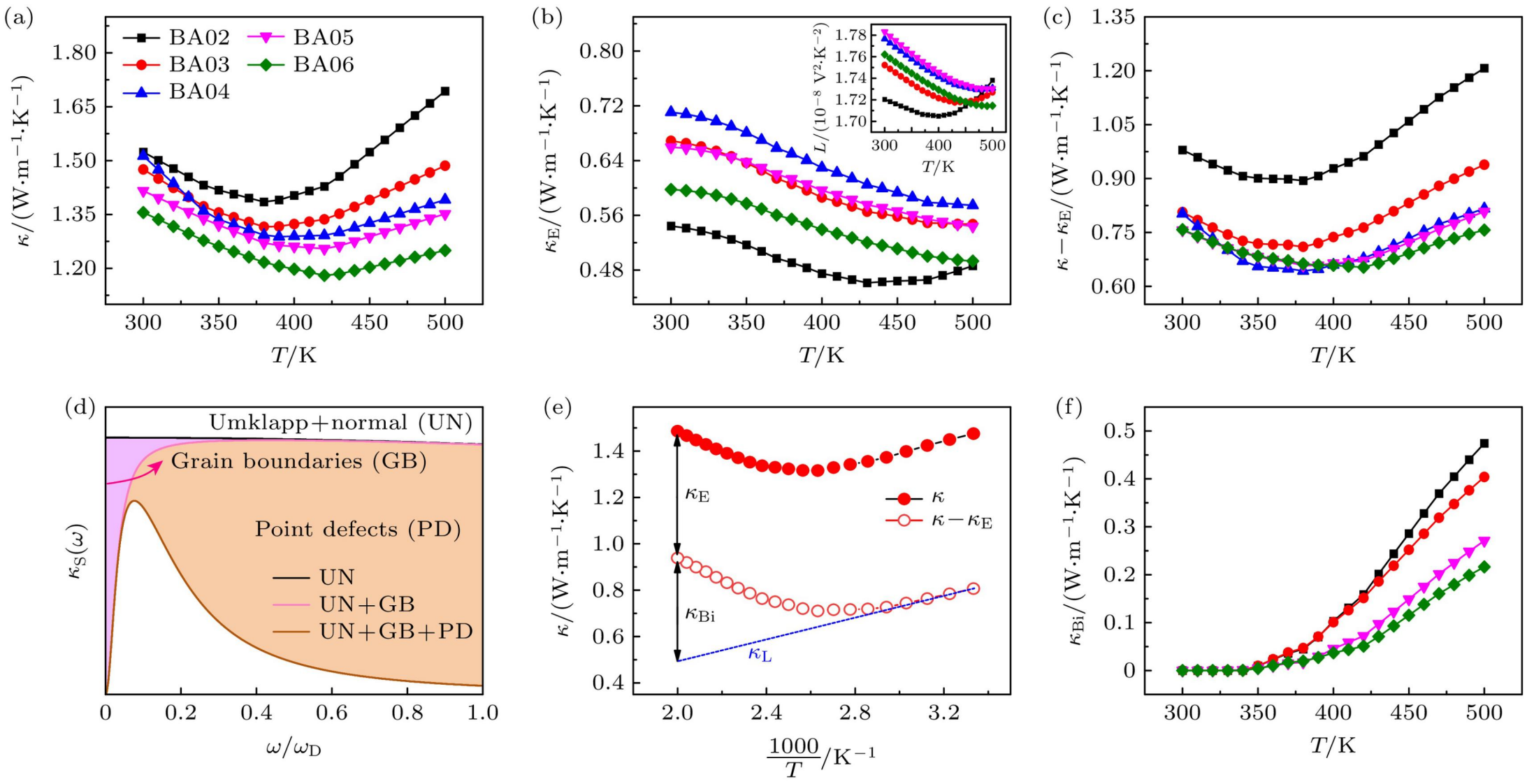

**Fig. 5** Temperature dependence of (a) total thermal conductivity, (b) carrier thermal conductivity $\kappa_E$, (c) lattice thermal conductivity $\kappa$–$\kappa_E$ for $Bi_2Te_{3-x}Se_x$ samples ($x$ = 0.2, 0.3, 0.4, 0.5 and 0.6) before annealing in the temperature range of 300–500 K, the inset in panel (b) shows the temperature dependence of Lorentz number $L$; (d) frequency-dependent spectral lattice thermal conductivity; (e) $T^{-1}$ dependence of $\kappa$ and $\kappa$–$\kappa_E$ in the temperature range of 300–500 K for BA03 sample, the dotted part is the dependence of $\kappa$–$\kappa_E$ and $T^{-1}$ in the range of 300–340 K; (f) temperature dependence of the bipolar thermal conductivity $\kappa_{Bi}$ for $Bi_2Te_{3-x}Se_x$ samples before annealing in the temperature range of 300–500 K.

$$L = \left(\frac{k_B}{e}\right)^2\left\{\frac{(r+7/2)F_{r+5/2}(\eta)}{(r+3/2)F_{r+1/2}(\eta)} - \left[\frac{(r+5/2)F_{r+3/2}(\eta)}{(r+3/2)F_{r+1/2}(\eta)}\right]^2\right\}. \quad (9)$$

Within the temperature range of 300–500 K, the electronic thermal conductivity ($\kappa_E$) of $Bi_2Te_{3-x}Se_x$ samples initially increases and then decreases as $x$ increases (Figure 5(b)). This trend aligns with the variation in electrical conductivity ($\sigma$) with Se doping. At high temperatures, the $\kappa_E$ of the sample with $x = 0.2$ increases significantly with rising temperature due to a substantial increase in the Lorenz number ($L$) (inset of Figure 5(b)). This phenomenon originates from the bipolar effect induced by intrinsic excitation at high temperatures. For samples with $x \geq 0.3$, $\kappa_E$ gradually decreases as temperature rises, consistent with the trend in $L$, without showing a significant increase at high temperatures. This behavior occurs because the increased band gap in heavily doped samples suppresses the bipolar effect generated by intrinsic excitation. At low temperatures, the concentration of minority carriers produced by intrinsic excitation is low, making their contribution to heat conduction negligible. Therefore, the lattice thermal conductivity ($\kappa_L$) can be calculated by subtracting $\kappa_E$ from the total thermal conductivity ($\kappa$). As shown in Figure 5(c), increasing the Se doping content facilitates the reduction of $\kappa_L$, particularly in the high-temperature region (400–500 K), where $\kappa_L$ decreases gradually with higher Se doping. The BA06 sample exhibits the lowest $\kappa_L$ of 0.65 W/(m·K) at 420 K. The significant reduction in $\kappa_L$ at high temperatures with increased Se doping may be attributed to two factors: 1) the difference in atomic radius and mass between Se and Te creates atomic-scale point defects when Se is doped into the $Bi_2Te_3$ lattice. These defects cause fluctuations in the mass and strain fields, thereby enhancing the scattering of heat-carrying phonons; 2) scattering at interfaces and grain boundaries generated by second phases. To better understand the relationship between $\kappa_L$ and various scattering mechanisms, we employed the Debye-Callaway model[43] to perform a quantitative analysis of the BA06 sample. Figure 5(d) illustrates the primary factors influencing phonon scattering at different frequencies.

$$\kappa_S(\omega) = \frac{k_B}{2\pi^2 v}\left(\frac{k_B T}{\hbar}\right)^3 \tau_{tot}(x)\frac{x^4 e^4}{(e^x-1)^2}, \tag{10}$$

In this equation, $\kappa_S(\omega)$ represents the frequency-dependent spectral lattice thermal conductivity; $\hbar$, $v$, and $\omega$ denote the reduced Planck constant, sound velocity, and phonon frequency, respectively; and $\tau_{tot}$ is the total relaxation time. The dimensionless variable $x = \hbar\omega/(k_B T)$ represents the reduced phonon frequency. Low-frequency phonons are primarily limited by grain boundary (GB) scattering, whereas high-frequency phonons with shorter wavelengths are effectively scattered by point defects (PD). This analysis indicates that the reduction in lattice thermal conductivity $\kappa_L$ is mainly attributed to the point defects introduced by Se doping. These point defects include $Se_{Te}^{\bullet}$, $Se_{Bi}^{'}$, and $Bi_{Te}^{'}$ antisite defects, as well as anion vacancies such as $V_{Te}^{\bullet\bullet}$.

Notably, in the temperature range of 420–500 K, the contribution of minority carriers to heat conduction gradually increases with rising temperature, leading to a significant increase in $\kappa - \kappa_E$. To elucidate the underlying mechanism, the contribution of bipolar conduction to $\kappa$ was also investigated. Since the effect of bipolar diffusion on heat conduction is negligible at low temperatures, $\kappa_L$ can be calculated using Eq. (11)[44]:

$$\kappa_{\mathrm{L}} = 3.5(\frac{\kappa_{\mathrm{Bi}}}{h})\frac{MV^{1/3}\theta_{\mathrm{D}}^{3}}{\gamma^{2}T}, \tag{11}$$

Here, $M$ represents the average atomic mass; $V$ denotes the average atomic volume; and $\theta_D$ and $\gamma$ correspond to the Debye temperature and the Grüneisen parameter, respectively. $\kappa_L$ exhibits a linear positive correlation with $T^{-1}$. Although $\kappa$–$\kappa_E$ at high temperatures gradually deviates from $\kappa_L$ due to the influence of bipolar thermal conductivity ($\kappa_{Bi}$), the high-temperature $\kappa_L$ of the sample can be obtained by extrapolating the low-temperature $\kappa_L$ data. Subsequently, the bipolar thermal conductivity is approximated by subtracting $\kappa_L$ and $\kappa_E$ from the total thermal conductivity $\kappa$ (Figure 5(e)), where the blue dashed line represents the high-temperature $\kappa_L$ calculated via extrapolation. Figure 5(f) illustrates the relationship between $\kappa_{Bi}$ and temperature for $Bi_2Te_{3-x}Se_x$ samples. Analysis indicates that increasing Se doping suppresses bipolar heat conduction. This occurs because higher Se doping widens the band gap, shifting the onset temperature of intrinsic excitation to higher values and thereby reducing bipolar thermal conductivity at high temperatures. As $x$ increases from 0.2 to 0.6, the contribution of the bipolar effect to the total thermal conductivity at 500 K decreases from 0.47 W/(m·K) to 0.20 W/(m·K), representing a 57% reduction. At low temperatures, intrinsic excitation is weak, and the contribution of minority carriers to heat conduction is negligible; thus, $\kappa$ can be approximated as the sum of $\kappa_E$ and $\kappa_L$. However, as temperature rises, the contribution of bipolar diffusion generated by intrinsic excitation to heat conduction becomes significant. In this regime, $\kappa$ comprises three components: $\kappa_E$, $\kappa_{Bi}$, and $\kappa_L$. With increasing Se doping, the influence of the bipolar effect on $Bi_2Te_{3-x}Se_x$ samples above 380 K diminishes continuously. Consequently, the $\kappa$ of the BA06 sample reaches a minimum of 1.18 W/(m·K) at 420 K, as shown in Figure 5(a).

The preceding analysis demonstrates that variations in Se doping significantly affect the high-temperature performance of the samples. To further investigate the underlying causes, we measured the ultraviolet-visible (UV-Vis) spectra of the $Bi_2Te_{3-x}Se_x$ samples, as presented in Figure 6(a). Within the tested wavelength range, only one distinct characteristic absorption peak was observed between 255 and 265 nm. No absorption peaks were detected in other wavelength ranges. Furthermore, as the Se doping level increased, particularly at a doping level of 0.6, a clear redshift (shift to longer wavelengths) of the absorption peak was evident. Using the UV-Vis spectral data and the Tauc plot (which characterizes the material's absorption edge), the optical direct band gap of the samples can be determined using the following relationship:

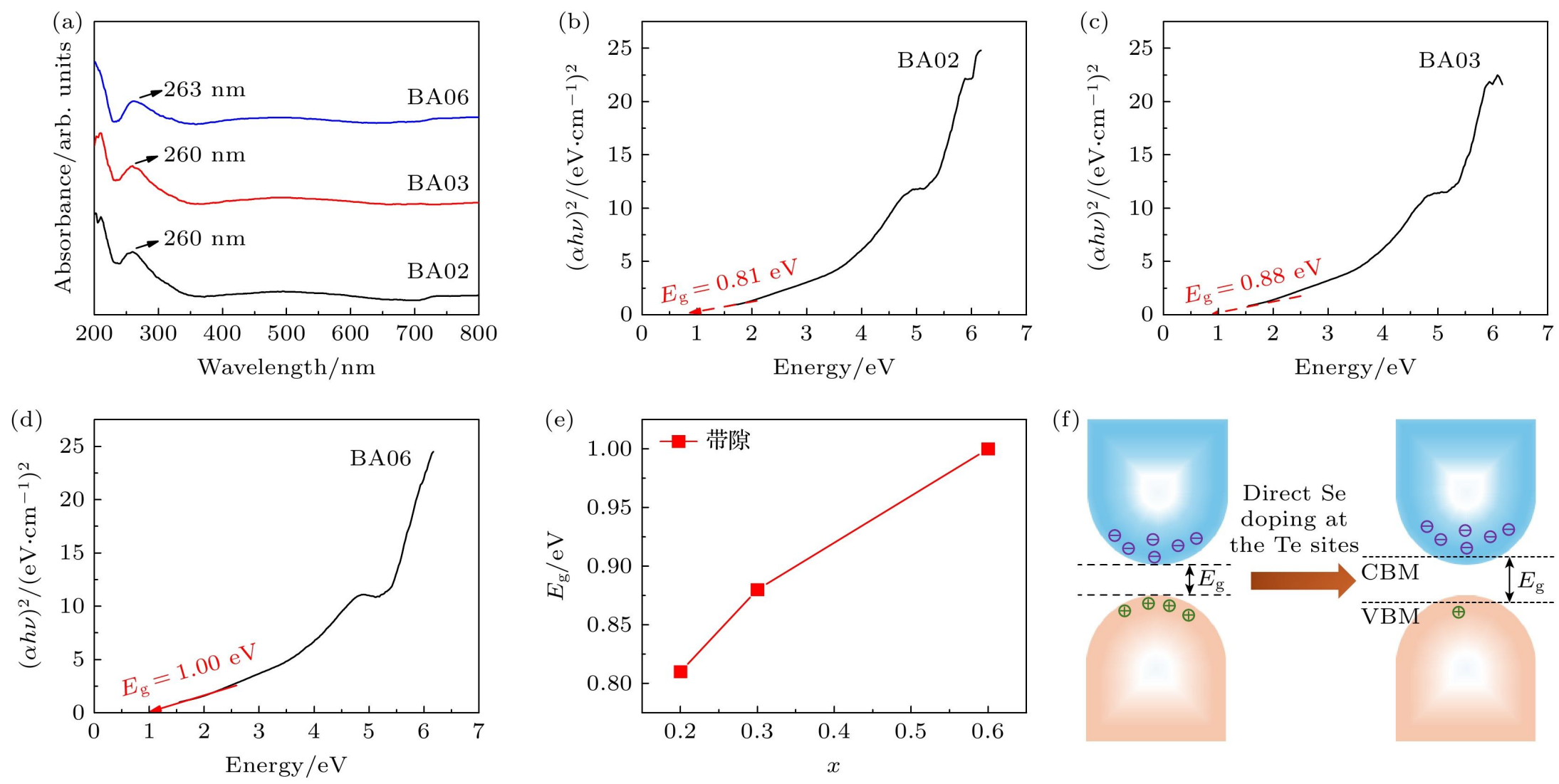

**Fig. 6** (a) Ultraviolet-visible (UV)-visible spectra; (b)–(d) plot of $(\alpha h\nu)^2$ versus $h\nu$ for direct optical bandgap of $Bi_2Te_{3-x}Se_x$ samples before annealing; (e) variation in the band gap for $Bi_2Te_{3-x}Se_x$ samples before annealing with the Se doping; (f) schematic of band gap enlargement by direct Se doping at the Te sites.

$$(\alpha h\nu)^2 = B(h\nu - E_g). \tag{12}$$

According to the optical absorption law, the relationship between the incident photon energy ($h\nu$) and the material absorption coefficient ($\alpha$) follows a specific functional form. Here, $B$ is a material-dependent proportionality constant, and $E_g$ represents the optical band gap. Based on the Tauc equation (Eq. (12)), the direct optical band gap of the material can be determined by plotting the function of $(\alpha h\nu)^2$ against $h\nu$. As $(\alpha h\nu)^2$ approaches zero, the corresponding photon energy $h\nu$ equals the direct band gap $E_g$. Consequently, the direct band gap values for each sample can be directly obtained from Fig. 6(b)–(d). The optical direct band gaps of the samples range from 0.81 to 1.00 eV. It is observed that the direct band gap increases with higher Se doping levels (Fig. 6(e)). This trend results from changes in phonon interactions and the shift of UV-Vis absorption peaks toward longer wavelengths due to varying Se doping concentrations[45]. The widening of the material band gap with increased Se doping (Fig. 6(f)) significantly weakens the high-temperature intrinsic excitation effect. The expanded band gap effectively reduces the thermal excitation concentration of carriers. This suppression mitigates the degradation of electrical and thermal transport properties in $Bi_2Te_{3-x}$ $Se_x$ samples at high temperatures (Fig. 4(b), Fig. 5(f)).

Figure 7 presents the thermal transport properties of $Bi_2Te_{3-x}$ $Se_x$ samples before and after annealing in the 300–500 K temperature range. As shown in Fig. 7(a), the total thermal conductivity decreases slightly for most samples after annealing, except for the $x = 0.2$ sample, which exhibits a minor increase. The change in carrier thermal conductivity before and after annealing correlates with the variation in electrical conductivity, as illustrated in Fig. 7(b). The lattice thermal conductivity decreases for all samples except the $x = 0.2$ sample, which shows a slight increase after annealing (Fig. 7(c)). This reduction stems from the decrease in the second phase Te and the formation of a small number of voids in heavily doped samples after annealing (Fig. 3). The AA06 sample exhibits a minimum $\kappa_L$ of 0.64 W/(m·K) at 420 K. The Lorenz number for all samples, calculated using Eq. (9), decreases (inset of Fig. 7(b)), which is consistent with the changes in the Seebeck coefficient.

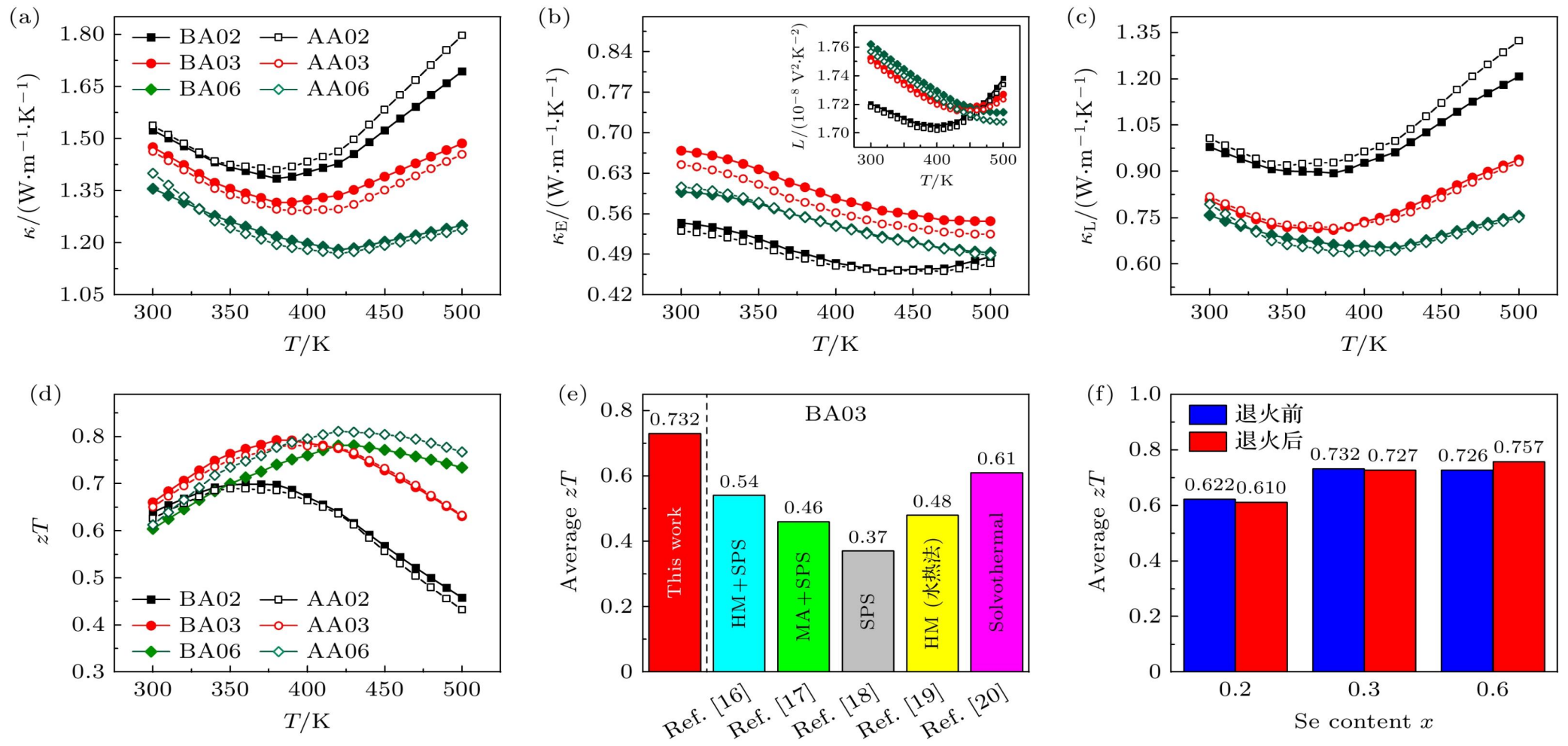

**Fig. 7** Temperature dependences of (a) total thermal conductivity, (b) carrier thermal conductivity $\kappa_E$, (c) lattice thermal conductivity $\kappa_L$ for $Bi_2Te_{3-x}Se_x$ samples before and after annealing in the temperature range of 300–500 K. The inset in panel (b) shows the temperature dependence of Lorentz number $L$. (d) Temperature dependence of $zT$ values for $Bi_2Te_{3-x}$ $Se_x$ alloys before and after annealing in the temperature range of 300–500 K. (e) Comparison of $zT_{ave}$ values (300–500 K) between optimized $Bi_2Te_{2.7}Se_{0.3}$ sample (before annealing) in this work and some reported Se-doped $Bi_2Te_{2-x}Se_x$ materials[16–20]. (f) $zT_{ave}$ values of samples with different Se doping amounts in the range of 300–500 K before and after annealing.

Figure 7(d) illustrates the temperature dependence of the $zT$ value for $Bi_2Te_{3-x}Se_x$ samples before and after annealing in the 300–500 K range. The $zT$ values of all samples initially increase and then decrease with rising temperature. The reduction in $zT$ at high temperatures is attributed to performance degradation caused by intrinsic excitation. In the 300–400 K range, the $zT$ value of $Bi_2Te_{3-x}Se_x$ samples first increases and then decreases as $x$ increases. Above 400 K, the $zT$ value gradually increases with increasing $x$. The BA03 sample achieves a maximum $zT$ value of 0.79 at 380 K. Meanwhile, the average $zT$ value ($zT_{ave}$) of $Bi_2Te_{3-x}Se_x$ samples in the 300–500 K range exhibits a similar trend with increasing $x$. The BA03 sample displays the highest $zT_{ave}$ of 0.73. This value exceeds those reported for some Se-doped n-type $Bi_2Te_{2-x}Se_x$ materials[16-20], as shown in Figure 7(e). Appropriate Se doping regulates carrier and point defect concentrations. This process increases electrical conductivity while reducing lattice thermal conductivity. Compared with other samples, the sample with $x$ = 0.3 exhibits optimal electrical transport properties and lower lattice thermal conductivity. Therefore, moderate Se doping synergistically optimizes the electrical and thermal transport properties of n-type $Bi_2Te_{3-x}Se_x$ materials. As shown in Figure 7(d), the $zT$ value of annealed samples improves significantly when the Se doping level exceeds 0.3. Before annealing, the sample with $x$ = 0.3 reached a maximum $zT$ of 0.79 at 380 K. After annealing, the sample with $x$ = 0.6 achieved a maximum $zT$ of 0.81 at 420 K. Furthermore, the $zT_{ave}$ in the 300–500 K range effectively increased from 0.73 ($x$ = 0.3) to 0.76 ($x$ = 0.6), as depicted in Figure 7(f). Annealing significantly enhances the performance of heavily doped samples. This improvement primarily stems from the optimization of the sample microstructure. These microstructural optimizations include a slight increase in grain size and crystallinity, a reduction in secondary phases, more uniform element distribution, and the emergence of a small number of micropores (Figures 2 and 3). In summary, the thermoelectric performance of $Bi_2Te_{3-x}Se_x$-based materials can be further optimized through annealing.

# 4 Conclusion

N-type $Bi_2Te_{3-x}Se_x$-based thermoelectric materials were successfully prepared using high-temperature melting combined with spark plasma sintering (SPS) and annealing processes. Moderate direct Se doping at Te sites synergistically optimizes electrical and thermal transport properties by regulating the defect structure, microstructure, and band gap of $Bi_2Te_{3-x}Se_x$ samples. The conclusions are as follows:

1) Controlling the Se doping level adjusts the concentration of $Bi_{Te}^{'}$ antisite defects and anion vacancies $V_{Te}^{\bullet\bullet}$. This regulation modulates carrier concentration and mobility, significantly increasing electrical conductivity and optimizing the electrical transport properties of $Bi_2Te_3$-based alloys.

2) Point defects arising from differences in atomic mass and radius between Se and Te atoms cause mass field and strain field fluctuations. These fluctuations strongly scatter high-frequency phonons, thereby significantly reducing lattice thermal conductivity.

3) Increasing the Se doping concentration widens the band gap of the material. This effectively suppresses intrinsic excitation at high temperatures and broadens the temperature range corresponding to optimal $zT$ values. The sample with $x$ = 0.3 achieves a maximum $zT$ of 0.79 at 380 K, with a peak $zT_{ave}$ of 0.73 in the 300-500 K range.

4) After annealing, $Bi_2Te_{3-x}$ $Se_x$ samples exhibit slight grain growth, improved

crystallinity, more uniform composition, reduced secondary phases, and the presence of a small number of micropores. These changes effectively regulate carrier transport and optimize electrical and thermal transport properties. The annealed sample with $x = 0.6$ achieves a maximum $zT$ of 0.81 at 420 K, with a peak $zT_{ave}$ of 0.76 in the 300-500 K range.

This study demonstrates that direct Se doping at Te sites, combined with annealing, effectively regulates carrier transport in $Bi_2Te_{3-x}Se_x$ samples. This approach modifies the defect structure, microstructure, and band gap. It suppresses the bipolar effect caused by intrinsic excitation and enhances phonon scattering. Consequently, it synergistically optimizes the electrical and thermal transport properties of the material. This process yields high-performance n-type $Bi_2Te_3$-based thermoelectric materials over a wide temperature range. This work provides insights for developing high-performance near-room-temperature thermoelectric materials suitable for broad temperature applications.